\documentstyle[aps,epsf]{revtex}  

\begin{document}        

\baselineskip 14pt
\title{Observation of the decay $K^+\to \pi^+ \nu \bar\nu$}
\author{Milind V. Diwan}
\address{Physics Department, Brookhaven National Laboratory, Upton, NY}
%
\maketitle              

\begin{abstract}        

We have observed 1 event consistent with the signature expected of 
the rare decay of a positive kaon to a positive pion and a 
neutrino anti-neutrino pair. In the examined momentum region of 
211 to 230 MeV/c in the center of mass of the kaon we estimated 
the backgrounds to be about $0.08 \pm 0.03$ events.
From this observation we estimate the branching ratio to be 
$4.2^{+9.7}_{-3.5} \times 10^{-10}$.  In this presentation I will explain the experiment, 
and the analysis techniques.  I will also discuss the expected 
improvements in the near future from the analysis of new data sets. 

\
\end{abstract}   	

\section{Introduction}               

The decay $K^+ \to \pi^+ \nu \bar\nu$~ has attracted interest due to
its sensitivity to $|V_{td}|$, the coupling of top to down quarks in
the Cabibbo-Kobayashi-Maskawa quark mixing matrix.  Theoretical
uncertainty in the branching ratio is minimal because the decay rate
depends on short distance physics and because the hadronic matrix
element can be extracted from the well-measured decay $K^+\to\pi^0
e^+\nu$.  After next-to-leading-logarithmic analysis of QCD
effects\cite{BB3}, calculation of isospin breaking, phase space
differences and other small corrections to the hadronic matrix
element~\cite{2mar}, and calculation of two-electroweak-loop
effects~\cite{twoloop}, the intrinsic uncertainty is only about
7\%\cite{bf}.  Based on current knowledge of Standard Model (SM)
parameters, the branching ratio $B(K^+\to\pi^+\nu\bar\nu)$ is expected
to be in the range $0.6 - 1.5 \times 10^{-10}$\cite{wdbll}.
Long-distance contributions to the branching ratio ({\it{i.e.}}
meson, photon exchange) appear to be negligible
($10^{-13}$)\cite{rhl,longd}.  
Since $K^+ \to \pi^+ \nu \bar\nu$~ is a
flavor changing neutral current process that is highly suppressed in
the SM, it also serves as a hunting ground for non-SM physics.  The
signature $K^+ \to \pi^+ $ `nothing'\cite{rhl,BSM,familon} 
includes
$K^+ \to \pi^+ \nu \bar \nu$ with non-SM intermediate states (such as
virtual supersymmetric particles), $K^+ \to \pi^+ \nu \bar \nu'$ (a
lepton flavor violating final state), $K^+ \to \pi^+ X^0 X^{0'}$ where
$X^0$ and $X^{0'}$ are not neutrinos, and $K^+ \to \pi^+ X^0$ where
$X^0$ is a single, non-interacting particle. Initial results from the
E787\cite{e787} experiment\cite{e787nim} at the Alternating Gradient Synchrotron
(AGS) of Brookhaven National Laboratory gave 90\% confidence level
(CL) upper limits $B$($K^+ \to \pi^+ \nu \bar\nu$) $< 2.4 \times
10^{-9}$ and $B$($K^+ \to \pi^+ X^0$)$< 5.2 \times 10^{-10}$ for a
massless $X^0$\cite{pnn96}. 
Here report on the analysis
of a new data sample with 2.4 times greater sensitivity, taken in 1995
using an upgraded beam and detector.

\section{The experiment}

 The signature for $K^+ \! \rightarrow \! \pi^+ \nu \overline{\nu}$ is
a $K^+$ decay to a $\pi^+$ of momentum $P<227$ MeV/$c$ and no other
observable product. Definitive observation of this signal requires
suppression of all backgrounds to well below the sensitivity for the
signal and reliable estimates of the residual background levels.
Major background sources include the copious two-body decays $K^+ \!
\rightarrow \! \mu^+ \nu_\mu$ ($K_{\mu 2}$) with a 64\% branching
ratio and $P=236$ MeV/$c$ and $K^+ \!  \rightarrow \!  \pi^+ \pi^0$
($K_{\pi 2}$) with a 21\% branching ratio and $P=205$ MeV/$c$. The
only other important background sources are scattering of pions in the
beam and $K^+$ charge exchange (CEX) reactions resulting in decays
$K_L^0\to\pi^+ l^- \overline\nu$, where $l=e$ or $\mu$.  To suppress
the backgrounds, we used the  redundant
kinematic and particle identification measurements and efficient
elimination of events with additional particles.

\begin{figure}[t]	
\centerline{\epsfxsize 6.5 truein \epsfbox{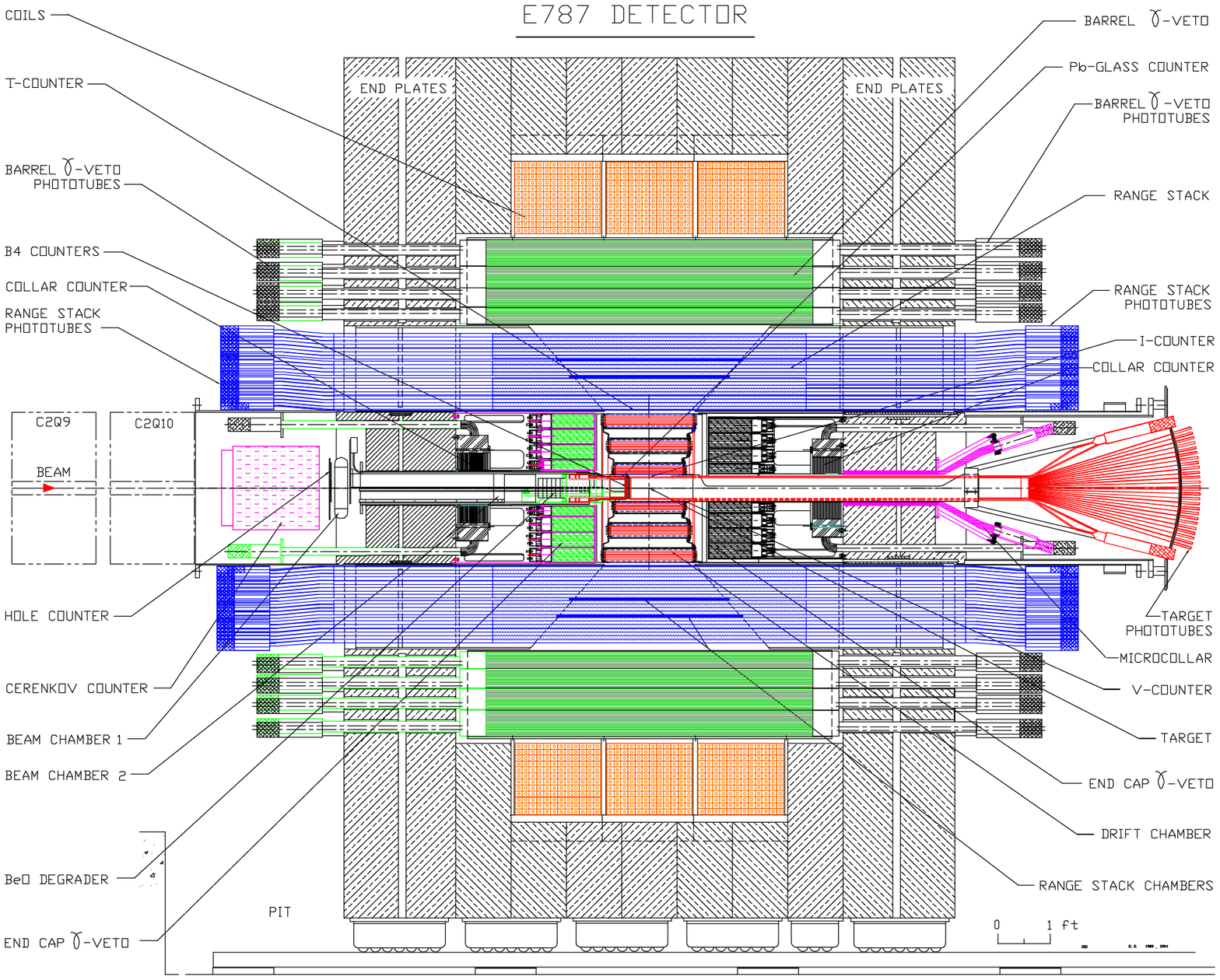}}   
\caption[]{
\label{detector}
\small Drawing of the E787 detector.}
\end{figure}

Kaons of 790~MeV/$c$~ were delivered to the experiment at a rate of
$7\times10^6$ per 1.6-s spill of the AGS. The kaon beam line (LESB3)
incorporated two stages of particle separation resulting in a pion
contamination of about 25\%.  The kaons were detected and identified
by \v{C}erenkov, tracking, and energy loss ($dE/dx$) counters.  About
20\% of the kaons slowed through a  degrader  to reach a stopping 
target of 5-mm-square plastic scintillating fibers read out by
500-MHz CCD transient digitizers\cite{ccd}.  Measurements of the
momentum ($P$), range ($R$, in equivalent cm of scintillator) and
kinetic energy ($E$) of charged decay products were made using the
target, a central drift chamber\cite{utc}, and a cylindrical range
stack with 21 layers of plastic scintillator and two layers of straw
tube tracking chambers. Pions were distinguished from muons by
kinematics and by observing the $\pi \!  \rightarrow \! \mu \!
\rightarrow \!  e$ decay sequence in the range stack using 500-MHz
flash-ADC transient digitizers (TD)\cite{tds}.  Photons were detected
in a $4\pi$-sr calorimeter consisting of a
14-radiation-length-thick barrel detector made of lead/scintillator
and 13.5 radiation lengths of undoped CsI crystal detectors (also read
out using CCD digitizers) covering each end\cite{csi}. In addition,
photon detectors were installed in the extreme forward and backward
regions, including a Pb-glass \v{C}erenkov detector just upstream of
the target.  A 1-T solenoidal magnetic field was imposed on the
detector for the momentum measurements.

\section{The analysis}

In the search for $K^+ \to \pi^+ \nu \bar\nu$, we required an
identified $K^+$ to stop in the target followed, after a delay of at
least 2 ns, by a single charged-particle track that was unaccompanied
by any other decay product or beam particle. This particle must
have been identified as a $\pi^+$ with $P$, $R$ and $E$ between the
$K_{\pi 2}$ and $K_{\mu 2}$ peaks.  A multilevel trigger selected
events with these characteristics for recording, and off-line analysis
further refined the suppression of backgrounds. To elude rejection,
$K_{\mu 2}$ and $K_{\pi 2}$ events would have to have been
reconstructed incorrectly in $P$, $R$ and $E$. In addition, any event
with a muon would have to have had its track misidentified as a pion
--- the most effective weapon here was the measurement of the $\pi \!
\rightarrow \!  \mu \!  \rightarrow \!  e$ decay sequence which
provided a suppression factor $10^{-5}$.  Events with photons, such as
$K_{\pi 2}$ decays, were efficiently eliminated by exploiting the full
calorimeter coverage. The inefficiency for detecting events with
$\pi^0$s was $10^{-6}$ for a photon energy threshold of about 1 MeV. A
scattered beam pion could have survived the analysis only by
misidentification as a $K^+$ and if the track were mismeasured as
delayed, or if the track were missed entirely by the beam counters
after a valid $K^+$ stopped in the target.  CEX background events
could have survived only if the $K_L^0$ were produced at low enough
energy to remain in the target for at least 2 ns, if there were no
visible gap between the beam track and the observed $\pi^+$ track, and
if the additional charged lepton went unobserved.

The data were analyzed with the goal of reducing the total expected
background to significantly less than one event in the final sample.
In developing the required rejection criteria (cuts), we took
advantage of redundant independent constraints available on each
source of background to establish two independent sets of cuts.  One
set of cuts was relaxed or inverted to enhance the background (by up
to three orders of magnitude) so that the other group could be
evaluated to determine its power for rejection.  For example, $K_{\mu
2}$ (including $K^+ \! \rightarrow \!  \mu^+ \nu_\mu \gamma$) was
studied by separately measuring the rejections of the TD particle
identification and kinematic cuts.  The background from $K_{\pi 2}$
was evaluated by separately measuring the rejections of the photon
detection system and kinematic cuts.  The background from beam pion
scattering was evaluated by separately measuring the rejections of the
beam counter and timing cuts.  Measurements of $K^+$ charge exchange
in the target were performed, which, used as input to Monte Carlo
studies, allowed the background to be determined.  Small correlations
in the separate groups of cuts were investigated for each background
source and corrected for if they existed.

The background levels anticipated with the final analysis cuts were
$b_{{K_{\mu 2}}} = 0.02 \pm 0.02$, $b_{{K_{\pi 2}}} = 0.03 \pm 0.02$,
$b_{Beam} = 0.02\pm 0.01$ and $b_{CEX} = 0.01 \pm 0.01$.  In total,
$b=0.08\pm 0.03$ background events were expected in the signal
region. 
This represents  an order of magnitude improvement in background
suppression  relative to ref.\cite{pnn96}, mainly because of 
improved kinematic and timing resolutions.
  Further confidence in the background estimates
and in the measurements of the background distributions near the
signal region was provided by extending the method described above to
estimate the number of events expected to appear when the cuts were
relaxed in predetermined ways so as to allow orders of magnitude
higher levels of all background types.  Confronting these estimates
with measurements from the full $K^+ \to \pi^+ \nu \bar\nu$~ data,
where the two sets of cuts for each background type were relaxed
simultaneously, tested the independence of the two sets of cuts.  At
approximately the $20 \times b$ level we observed 2 events where $1.6
\pm 0.6$ were expected, and at the level $150 \times b$ we found 15
events where $12 \pm 5$ were expected.  Under detailed examination,
the events admitted
by the relaxed cuts
were consistent with
being due to the known background sources.  Within the final signal
region, we still had additional background rejection capability.
Therefore, prior to looking in the signal region, we established
several sets of ever-tighter criteria which were designed to be used
only to interpret any events that fell into the signal region.

\begin{figure}
\begin{minipage}{0.24\linewidth}
\flushright{\epsfxsize 2.62 truein \epsfbox{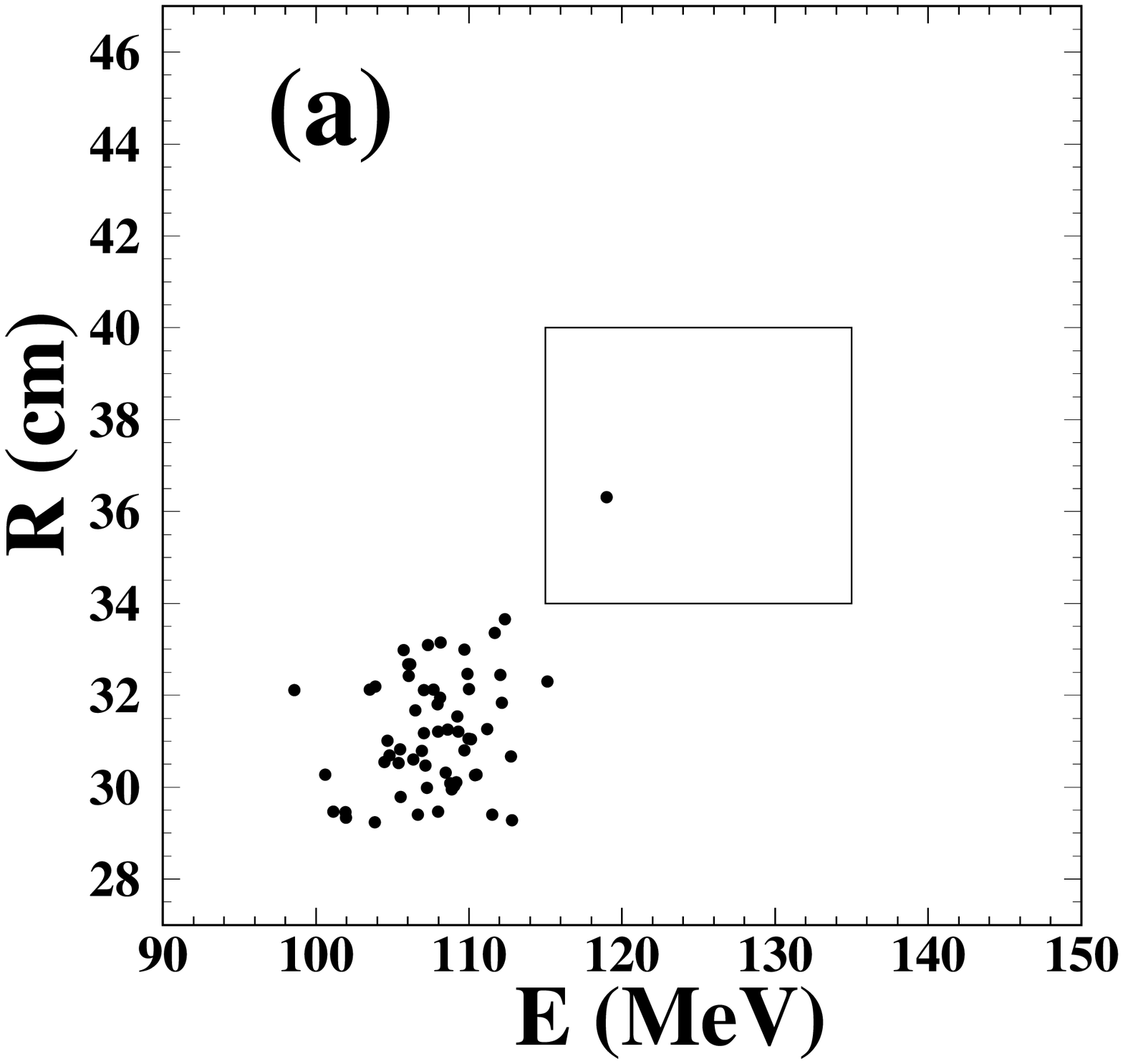}}
\end{minipage}\hfill
\begin{minipage}{0.24\linewidth}
\flushleft{\epsfxsize 2.62 truein \epsfbox{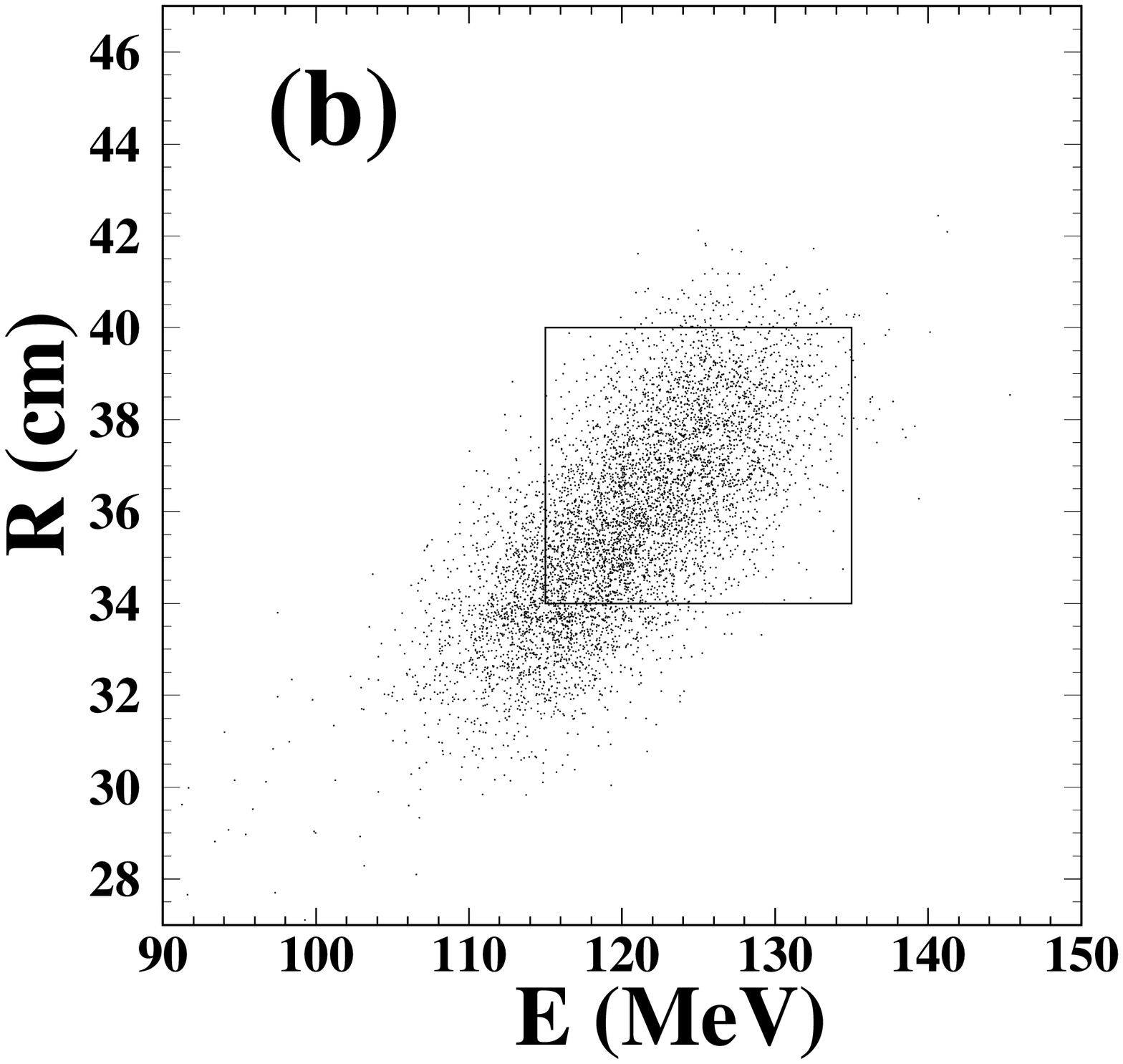}}
\end{minipage}\hfill
\caption{\label{data2}}{(a) Range ($R$) vs. energy ($E$) distribution
for the $K^+ \to \pi^+ \nu \bar\nu$~ data set with the final cuts
applied. The box enclosing the signal region contains a single
candidate event. (b) The Monte Carlo simulation of $K^+ \to \pi^+ \nu
\bar\nu$~ with the same cuts applied.}
\end{figure}

Figure~\ref{data2}(a) shows $R$ vs.~$E$ for the events surviving all
other analysis cuts. Only events with measured momentum in the
accepted region $211 \le P \le 230$ MeV/$c$ are plotted. The
rectangular box indicates the signal region specified as range $34 \le
R \le 40$ cm of scintillator (corresponding to $214 \le P_{\pi} \le
231$ MeV/$c$) and energy $115 \le E \le 135$ MeV ($213 \le P_{\pi} \le
236$ MeV/$c$) which encloses the upper 16.2\% of the $K^+ \to \pi^+
\nu \bar\nu$~ phase space.  One event was observed in the signal
region. The residual events below the signal region clustered at $E=
108$ MeV were due to $K_{\pi 2}$\ decays where both photons had been
missed. The number of these events is consistent with estimates of the
photon detection inefficiency.

\begin{figure}
\leavevmode 
\centerline{\epsfxsize 5.6 in \epsfbox{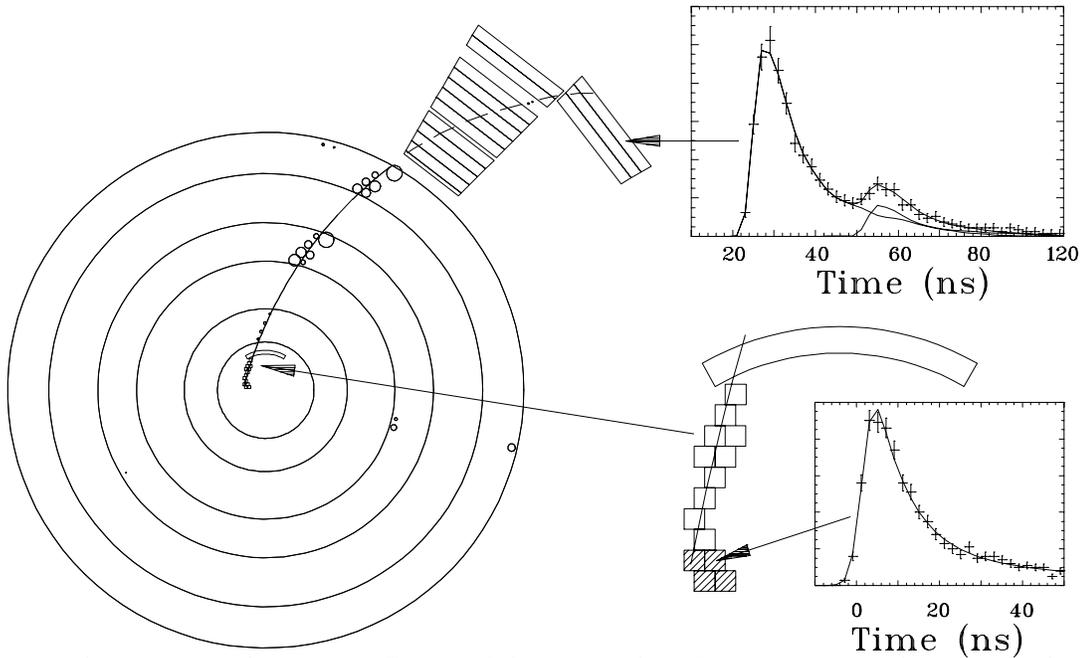}}
\caption{\label{event1}Reconstruction of the candidate event. On the
left is the end view of the detector showing the track in the target,
drift chamber (indicated by drift-time circles), and range stack
(indicated by the layers that were hit). At the lower right is a
blowup of the target region where the hatched boxes are kaon hits, the
open boxes are pion hits, and the inner trigger counter hit is also
shown.  The pulse data sampled every 2 ns (crosses), in one of the
target fibers hit by the stopped kaon is displayed along with a fit
(curve) to the expected pulse shape. At the upper right of the figure
is the $\pi \to \mu$ decay signal in the range stack scintillator
layer where the pion stopped.  The crosses are the pulse data sampled
every 2 ns, and the curves are fits for the first, second and combined
pulses.}
\end{figure}

A reconstruction of the candidate event is shown in Fig.~\ref{event1}.
Measured parameters of the event include $P=219.1\pm2.9$ MeV/$c$,
$E=118.9\pm3.9$ MeV, $R=36.3\pm 1.4$ cm, and decay times $K\to\pi$,
$\pi \to \mu$ and $\mu \to e$ of $23.9\pm0.5$ ns, $27.0\pm0.5$ ns and
$3201.1\pm 0.7$ ns, respectively.  No significant energy was observed
elsewhere in the detector in coincidence with the pion.
The event also satisfied the most demanding criteria designed in
advance for candidate evaluation.  This put it in a region with an
additional background rejection factor of 10. In this region,
$b'=0.008 \pm 0.005$ events would be expected from known background
sources while 55\% of the final acceptance for $K^+ \to \pi^+ \nu
\bar\nu$~ would be retained\cite{bclean}.  Since the explanation of
the observed event as background is highly improbable, we conclude
that we have likely observed a kaon decay $K^+ \to \pi^+ \nu \bar\nu$.

To calculate the branching ratio indicated by this observation, we
used the final acceptance for $K^+ \to \pi^+ \nu \bar\nu$, $A=0.0016
\pm 0.0001^{stat} \pm 0.0002^{syst}$, and the total exposure of 
$N_{K^+}=1.49\times
10^{12}$ kaons entering the target.  Where possible, we employed
calibration data taken simultaneously with the physics data for the
acceptance calculation. We relied on Monte Carlo studies only for the
solid angle acceptance factor, the $\pi^+$ phase space factor and the
losses from $\pi^+$ nuclear interactions and decays in flight. The details 
of these calculations are in various Ph.D. theses \cite{jesse}.
 If the
observed event is due to $K^+ \to \pi^+ \nu \bar\nu$, the branching
ratio is $B(K^+ \! \rightarrow \!  \pi^+ \nu \overline{\nu}) =
4.2^{+9.7}_{-3.5} \times 10^{-10}$.

The likelihood of the candidate event being due to $K^+\to \pi^+ X^0$
($M_{X^0} = 0$) is small. Based on the measured resolutions, the
$\chi^2$ CL for consistency with this hypothesis is 0.8\%.  Thus,
using the acceptance for $K^+\to \pi^+ X^0$, $A_{(K^+ \to \pi^+
X^0)}=0.0052 \pm 0.0003^{stat} \pm 0.0007^{syst}$, and no observed
events in the region $221 < P < 230$ MeV/$c$, a 90\% CL upper limit of
$B(K^+ \!  \rightarrow \!  \pi^+ X^0) < 3.0 \times 10^{-10}$ was
derived.

The observation of an event with the signature of $K^+ \to \pi^+ \nu
\bar\nu$~ is consistent with the expectations of the SM which are
centered at about $1 \times 10^{-10}$. Using the result for $B$($K^+
\to \pi^+ \nu \bar\nu$) and the relations given in ref.~\cite{BB3},
$|V_{td}|$ lies in the range $0.006<|V_{td}|< 0.06$.  E787
has recently collected additional data and the experiment is
continuing.

\section{Future expectations}

The E787 experiment has had four runs during 1995--98.
The typical conditions for the 1995 run were $13\times 10^{12}$ protons per 
AGS spill, 5.3 MHz of
incident $K^+$ of 790 MeV/c, a stopped kaon rate of 1.2 M/spill, a deadtime of
25\%, and an acceptance of 0.16\%. Over the course of the years we
steadily increased the duty factor of the AGS from 41 to 52 percent.
 We also reduced the momentum 
of the kaons to 710 MeV/c to increase the fraction that stop in the detector;
this lowered the accidental rates in the detector. 
 The expected sensitivity from the 1995--98 runs is $\sim\times$4.4  that
of the 1995 data alone
without considering potential improvements in the analysis.
A preliminary re-analysis of the E787 1995
data with improvements in the analysis software have demonstrated a
background rejection that is $\sim\times$2.3 larger.
This background level (roughly equivalent to a branching
ratio of $1.5\times10^{-11}$) is sufficient for future measurements
of the $K^+\to \pi^+ \nu \bar\nu$
 branching ratio.  Results of the analysis of the larger
data set are expected within a few months.

A new experiment,
E949, recently received approval and is
expected to run at the AGS starting
in the year 2001. This experiment is designed to reach a sensitivity
of (8--14)$\times10^{-12}$, an order of magnitude below the Standard
Model prediction and to determine $|V_{td}|$ to better than 27\%. It is
built around the existing E787 detector to take advantage of the
extensive analysis of that detector, allowing a reliable projection of
the new experiment to the required sensitivity with a high level of
confidence.

The E949 detector will have significantly upgraded photon veto
systems, data acquisition 
and trigger compared to the E787 experiment. 
The photon veto  upgrade includes a barrel veto liner that will replace
the outer layers of the range stack. 
It is 2.3 $X_\circ$ thick and will add substantially to the
thin region at 45$^\circ$. Additional photon veto upgrades will be installed
along the beam direction. The most important data acquisition upgrade will
be to instrument the range stack  with TDC's to extend the search time for
the Michel electron ($\mu^+\rightarrow e^+$) and to allow the
transient digitizer range to be shortened. The shortening of the 
transient digitizer range should
allow a reduction of deadtime by 30--50\%. Trigger upgrades
should reduce the deadtime further and reduce the acceptance
loss due to the online photon veto.
Compared to the E787 running conditions in 1995 an improvement
of 50\% has already been realized. Additional improvements in
these areas and in offline software are expected to gain another
90\%. Additional sensitivity gains can be realized by 
including the region of phase space below the K$_{\pi2}$ peak and
by reoptimizing the analysis algorithms to run at higher rates.
Each of these should provide a factor of 2 more sensitivity.
The total
gain in sensitivity per hour will be 6--13 times over the E787
published result on the 1995 data set.

\section{Conclusion}

\begin{figure}[tb]
\centerline{\epsfxsize 4 truein \epsfbox{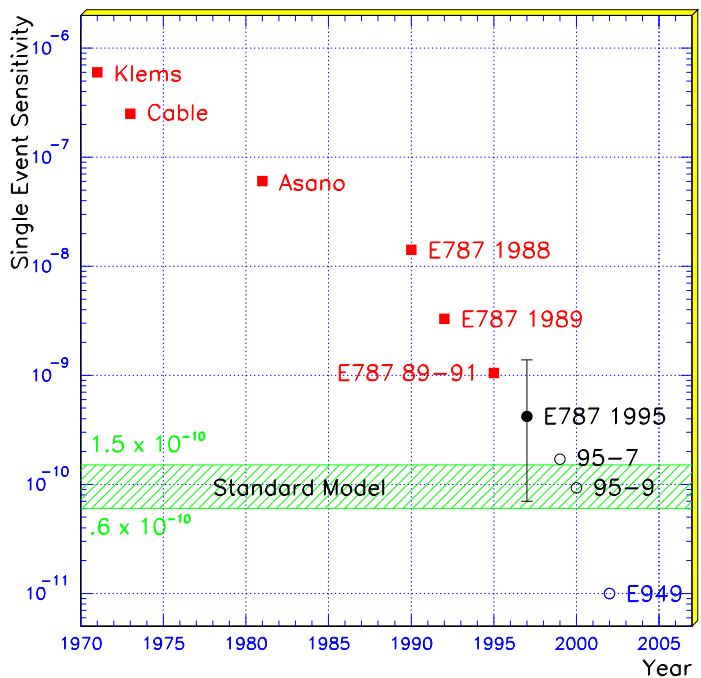}}
\caption{ \label{figbrt} History of progress in the search for $K^+\to 
\pi^+ \nu \bar\nu$.
The sensitivity of experiments setting limits is shown in
solid squares. The first actual measurement is shown as a solid circle and the projected
future measurements are shown as open circles. The background levels are
shown as stars for the recent data.}
\end{figure}

The prospects for further improvement in the determination of B($K^+\to 
\pi^+ \nu \bar\nu$)
are bright. The first observation of this rare and interesting decay
has recently been published. The data on hand, or soon to be
available, from the E787 experiment, should provide almost an order of
magnitude more sensitivity. The recently approved experiment E949
should reach at least a factor of five further than E787 and make a
very interesting measurement of $|V_{td} |$.  There is also a proposal,
CKM, at the FNAL Main Injector, to push even further, to $10^{-12}$ by
looking for the decay in flight.  A plot showing the progress from
past, current and approved experiments  is shown in
Figure~\ref{figbrt}. 
The search for this decay, with its very clean and well understood
prediction within the standard model, could soon provide either a crucial
test of the standard model or a precise measurement of $|V_{td}|$. 

\begin{references}  

\bibitem{e787}
The E787 collaboration: S.~Adler, M.S.~Atiya, I-H.~Chiang, M.V.~Diwan,
  J.S.~Frank, J.S.~Haggerty, S.H.~Kettell, 
  T.F.~Kycia, K.K.~Li, L.S.~Littenberg, C.~Ng,
  A.~Sambamurti,  A.~Stevens,
  R.C.~Strand,C.~Witzig, T.K.~Komatsubara, M.~Kuriki,
  N.~Muramatsu, S.~Sugimoto, T.~Inagaki, S.~Kabe,
  M.~Kobayashi, Y.~ Kuno, T.~Sato, T.~Shinkawa,
  Y.~Yoshimura, Y.~Kishi, T.~Nakano,
  M.~Ardebili, A.O.~Bazarko, M.R.~Convery,
  M.M.~Ito, D.R.~Marlow,
  R.A.~McPherson, P.D.~Meyers, F.C.~Shoemaker,
  A.J.S.~Smith, J.R.~Stone, M.~Aoki,
  E.W.~Blackmore, P.C.~Bergbusch, D.A.~Bryman, 
A.~Konaka,
  J.A.~Macdonald, J.~Mildenberger, T.~Numao,
  P.~Padley, J.-M.~Poutissou,
R.~Poutissou,
  G.~Redlinger, J.~Roy,
  A.S.~Turcot,
P.~Kitching and
   R.~Soluk

\bibitem{BB3}
G. Buchalla and A.J. Buras,  Nucl. Phys. {\bf B412}, 106 (1994).

\bibitem{2mar} W.J.~Marciano~and~Z.~Parsa, Phys. Rev. D {\bf 53}, R1
(1996).

\bibitem{twoloop}
G. Buchalla and A.J. Buras, SLAC-PUB-7575, TUM-HEP-280/97, hep-ph/9707243.

\bibitem{bf} A.J. Buras and R. Fleischer,
TUM-HEP-275-97, hep-ph/9704376, {\it Heavy Flavours II}, World
Scientific, eds. A.J.Buras and M. Linder (1997), to be published.

\bibitem{wdbll} G. Buchalla, A.J. Buras and M.E. Lautenbacher,
Rev. Mod. Phys. {\bf 68}, 1125 (1996).

\bibitem{rhl} J.S.~Hagelin and L.S.~Littenberg,  Prog. Part. Nucl. Phys.
{\bf 23}, 1 (1989).

\bibitem{longd} D. Rein and L.M. Sehgal, Phys. Rev. D {\bf 39}, 3325 (1989);
M. Lu and M.B. Wise, Phys. Lett. {\bf B324}, 461 (1994);
C.Q.~Geng, I.J.~Hsu, and Y.C.~Lin,  Phys. Lett. {\bf B355}, 569 (1995);
S.~Faijfer, Nuovo. Cim. {\bf 110A}, 397 (1997).


\bibitem{BSM} Y.~Grossman~and~Y. Nir,  Phys.\ Lett. {\bf B398}, 163 (1997);
G.~Couture~and~H.~K$\ddot{\mbox{o}}$nig,  Z.\ Phys. {\bf C69}, 167 (1996);
I.I.~Bigi and F.~Gabbiani,  Nucl. Phys. {\bf B367}, 3 (1991);
K. Agashe and M. Graesser, Phys. Rev. D {\bf 54}, 4445 (1996);
M.~Leurer,  Phys.\ Rev.\ Lett. {\bf 71}, 1324 (1993);
S.~Davidson, D.~Bailey, and B.~Campbell,  Z.\ Phys. {\bf C61}, 613 (1994);
S.~Bertolini~and~A.~Santamaria,  Nucl.\ Phys. {\bf B315}, 558 (1989).

\bibitem{familon} F.~Wilczek,  Phys. Rev. Lett. {\bf 49}, 1549 (1982).

\bibitem{e787nim} M.S. Atiya  {\it{et al.}}, Nucl. Instr. Meth.
                  {\bf A321}, 129 (1992).
\bibitem{pnn96} S. Adler  {\it{et al.}},  Phys. Rev. Lett. {\bf 76}, 1421 
(1996).

\bibitem{ccd} D.A. Bryman  {\it{et al.}}, 
Nucl. Instr. Meth. (1997), to be published.
\bibitem{utc} E.W. Blackmore  {\it{et al.}}, 
Nucl. Instr. Meth. (1997), to be published.

\bibitem{tds} M. Atiya  {\it{et al.}}, Nucl. Instr. Meth.
                  {\bf A279}, 180 (1989).
\bibitem{csi} I-H. Chiang  {\it{et al.}}, IEEE
Trans. Nucl. Sci. {\bf NS-42}, 394 (1995).

\bibitem{jesse} J. R. Stone, Princeton University Ph. D. thesis, May 1998.


\bibitem{bclean} The background estimates in the region with the tightest
cuts were $b_{{K_{\mu 2}}} = 0.004 \pm 0.004$, $b_{{K_{\pi 2}}} =
0.001 \pm 0.001$, $b_{Beam} = 0.002\pm 0.002$ and $b_{CEX} = 0.002 \pm
0.002$.


 \end{references}
\end{document}